\documentclass[twocolumn]{revtex4} 
\usepackage{dcolumn}
\usepackage{bm}
\usepackage[utf8]{inputenc}
\usepackage[russian,ukrainian,english]{babel}
\usepackage{amssymb,theorem,xspace}
\usepackage{amsmath}
\usepackage{mathrsfs}
\usepackage{pgfplots}
\usepackage{graphicx}
\usepackage{scalerel,stackengine}
\stackMath
\newcommand\reallywidehat[1]{%
\savestack{\tmpbox}{\stretchto{%
  \scaleto{%
    \scalerel*[\widthof{\ensuremath{#1}}]{\kern-.6pt\bigwedge\kern-.6pt}%
    {\rule[-\textheight/2]{1ex}{\textheight}}
  }{\textheight}%
}{0.5ex}}%
\stackon[1pt]{#1}{\tmpbox}%
}

\newcommand{\newc}{\newcommand}
\newc{\eq}{equa\c{c}\~ao}
\newc{\eqs}{equações}
\newc{\Eq}{Equa\c{c}\~ao}
\newc{\Eqs}{equações}
\newc{\s}{s\~ao}
\newc{\n}{n\~ao}
\newc{\sch}{Schr\"{o}dinger}
\newc{\dkp}{Duffin-Kemmer-Petiau}
\newc{\se}{se\c{c}\~ao}
\newc{\fu}{fun\c{c}\~ao}
\newc{\kg}{Klein-Gordon}
\newc{\del}{\partial}
\newc{\rpdv}[1]{\frac{\overrightarrow{\partial}}{\partial{#1}}}
\newc{\lpdv}[1]{\frac{\overleftarrow{\partial}}{\partial{#1}}}
\newc{\dround}[2]{\del_{#1}\del^{#2}}
\newc{\at}[2][]{#1\bigg|_{#2}}
\begin{document}
\title{Symplectic Field Theory of the Galilean Covariant Scalar and Spinor Representations}%
\author{Petronilo, G.X.A.}
\affiliation{ Centro Internacional de F\'isica da Mat\'eria Condensada, Universidade de Bras\'ilia}
\address{Campus Universit\'ario Darcy Ribeiro, Bras\'ilia-DF | CEP 70910-900}
\email{gustavopetronilo@gmail.com}

\author{Ulhoa, S.}
\affiliation{ Centro Internacional de F\'isica da Mat\'eria Condensada, Universidade de Bras\'ilia}
\address{Campus Universit\'ario Darcy Ribeiro, Bras\'ilia-DF | CEP 70910-900 }

\author{Santana, A.E.}
\affiliation{ Centro Internacional de F\'isica da Mat\'eria Condensada, Universidade de Bras\'ilia}
\address{Campus Universit\'ario Darcy Ribeiro, Bras\'ilia-DF | CEP 70910-900 }


\setcounter{page}{1}%

\date{\today} 
\begin{abstract}
Exploring the concept of the extended Galilei group $\mathcal{G}$, a representation for the symplectic quantum mechanics in the manifold of $\mathcal{G}$, written in the light-cone of a five-dimensional de Sitter space-time,  is derived consistently with the method of the Wigner function. A Hilbert space is constructed endowed with a simplectic structure, studying unitary operators describing rotations and translations, whose generators satisfy the Lie algebra of $\mathcal {G}$. This representation gives rise to the {\sch} (Klein-Gordon-like) equation for the wave functions in phase-space, such that the dependent variables have the position and linear momentum contents. Wave functions are associated with the Wigner function through the Moyal product, such that the wave functions represent a quasi-amplitude of probability. We construct the Pauli-{\sch} (Dirac-like) equation in phase-space in its explicitly covariant form. Finally, we show the equivalence between the five dimensional formalism of phase-space with the usual formalism, proposing a solution that recover the non-covariant form of the Pauli-{\sch} equation in phase-space.
\end{abstract}
\maketitle

\vspace{2cm}
 \section{Introduction}

 In 1988 Takahashi \emph{et. al.} \cite{takahashi1988towards} began a study of Galilean covariance, where it was possible to develop an explicitly covariant non-relativistic field theory. With this formalism, the Schr\"{o}dinger equation takes a similar form as Klein-Gordon equation in the light-cone of a  (4,1) de Sitter Space~\cite{omote1989galilean, santana}. With the advent of the Galilean covariance, it was possible to derive  the non-relativistic version of the Dirac theory, which is known in its usual form as the Pauli-Schr\"{o}dinger equation. The goal in the present work is to derive a Wigner representation for such covariant theory. \vskip 2pt
Wigner quasi-probability distribution (also called the Wigner function or the Wigner-Ville distribution in honor of Eugene Wigner and Jean-Andr\'e Ville) was introduced by Eugene Wigner in 1932~\cite{wigner1932uberschreiten}, in order to study quantum corrections to classical statistical mechanics. The aim was to relate the wave function that appears in the Schr\"{o}dinger equation to a probability distribution in phase space. It is a generating function for all the spatial autocorrelation functions of a given quantum mechanical function $\psi(x)$. Thus, it maps the quantum density matrix into the real phase space functions and operators introduced by Hermann Weyl in 1927~\cite{weyl1927quantenmechanik}, in a context related to the theory of representation in mathematics (Weyl quantization in physics). Indeed, this is the Wigner-Weyl transformation of the density matrix; i.e. the realization of that operator in the phase space. It was later re-derived by Jean Ville in 1948 ~\cite{ville1948cables} as a quadratic representation (in sign) of the local time frequency energy of a signal, effectively a spectrogram. In 1949, Jos\'e Enrique Moyal ~\cite{moyal1949quantum}, who independently derived the Wigner function, as the functional generator of the quantum momentum,  as a basis for an elegant codification of all expected values and, therefore, of quantum mechanics in phase space formulation (phase space representation). This representation has been applied to a number of areas such as in statistical mechanics, quantum chemistry, quantum optics, classical optics and signal analysis in several fields, such as electrical engineering, seismology, time-frequency analysis for music signals, spectrograms in biology and speech processing and motor design. In order to derived a phase space representations for the Galilean-covariant spin 1/2 particles, we use a symplectic representation for the Galilei group, which is associated with the Wigner approach~\cite{oliveira2004symplectic, amorim, hara, rendisley2018}.

This article is divided as follows. In Section~\ref{Galilean Covariance} the construction of the Galilean Covariance is presented. The Schr\"{o}dinger ({Klein-Gordon}-like) equation and Pauli-Schr\"{o}dinger (Dirac-like) equation are derived showing the equivalence between our formalism with the usual non-relativistic formalism are demonstrated. In Section~\ref{Symplectic Quantum Mechanics and the Galilei Group} a symplectic structure is constructed in the Galilean manifold. Using the commutation relations, the Schr\"{o}dinger equation in five dimensions in phase space are constructed. With a proposed solution, the Schr\"{o}dinger {equation} in phase space is restored to its non-covariant form in (3+1) dimensions. The explicitly covariant Pauli-Schr\"{o}dinger {equation} is derived in Section~\ref{Spin 1/2 Symplectic Representaion}. We study the Galilean spin 1/2 particle with a external potential and solutions are proposed and discussed. In Section~\ref{Concluding Remarks}, final concluding remarks are presented.

\section{Galilean Covariance}\label{Galilean Covariance}
The Galilei transformations are given by
\begin{eqnarray}
      \boldsymbol{x}' = R\boldsymbol{x}+\boldsymbol{v}t+\boldsymbol{a}\label{tg-1}\\
      t^\prime = t+b\label{tg-2}
\end{eqnarray}
where $R$ stands for the three-dimensional Euclidian rotations, $v$ is the relative velocity defining the Galilean boosts, $\textbf{a}$ stands for spacial translations and $\textbf{b}$, for time translations.
Consider a free mass particle $m$; the mass shell relation is given by $\widehat{P}^2-2mE=0$.\vskip 2pt
We can then define a 5-vector, $p^\mu=(p_x,p_y,p_z,m,E)=(p^i,m,E)$, with $i=1,2,3$.
\vskip 2pt
Thus, we can define a scalar product of the type
\begin{eqnarray}
p_\mu p_\nu g^{\mu\nu}&=&p_ip_i-p_5p_4-p_4p_5=\widehat{P}^2-2mE=k,
\end{eqnarray}
where $g^{\mu\nu}$ is the metric of the space-time to be construct, e $p_\nu g^{\mu\nu}=p^\mu$.
\vskip 2pt
Let us define a set of canonical coordinates $q^\mu$ associated with $p^\mu$, by writing a five-vector in $M$, $q^\mu=(\textbf{q},q^4,q^5)$, $\textbf{q}$ is the canonical coordinate associated with $\widehat{P}$;
$q^4$ is the canonical coordinate associated with $E$, and thus can be considered as the time coordinate;
$q^5$ is the canonical coordinate associated with $m$
explicitly given in terms of \textbf{q} and $q^4$,
$q^\mu q_\mu=q^\mu q^\nu\eta_{\mu\nu}= \textbf{q}^2-2q^4q^5=s^2$. From this $q^5=\frac{\textbf{q}^2}{2t}$; or infinitesimally, we obtain $\delta q^5 =\textbf{ v}\cdot\delta\frac{\textbf{q}}{2}$. Therefore, the fifth component is basically defined by velocity.

That can be seen as a special case of a scalar product in $G$ denoted as
\begin{eqnarray}
    (x|y)=g^{\mu\nu}x_{\mu}y_{\nu} = \sum_{i=1}^3 x_{i}y_{i}-x_{4}y_{5}-x_{5}y_{4},
\end{eqnarray}
where $x^4=y^4=t$, $x^5=\frac{x^2}{2t}$ e $y^5=\frac{y^2}{2t}$. Hence, the following the metric can be introduced

   \begin{equation}\label{g}
   (g_{\mu\nu}) = \left(\begin{array}{ccccc}
   1&0&0&0&0\\
   0&1&0&0&0\\
   0&0&1&0&0\\
   0&0&0&0&-1\\
   0&0&0&-1&0
   \end{array}\right).
\end{equation}
This is the metric of Galilean manifold $\mathcal{G}$. In the sequence this structure is explored in order to study unitary representations.
\section{Hilbert Space and Sympletic Structure}
\vskip 2pt
Consider $\mathcal{G}$ an analytical manifold where each point is specified by coordinates $q_\mu$, with $\mu = 1, 2, 3,4,5$ and metric specified by \eqref{g}.
The coordinates of each point in the cotangent-bundle $T^{*}\mathcal{G}$ will be denoted by $(q_\mu, p_\mu)$. The space $T^{*}\mathcal{G}$ is equipped with a symplectic structure via a 2-form.
\begin{equation}
    \omega=dq^\mu\wedge dp_\mu
\end{equation}
called the symplectic form (sum over repeated indices is assumed). We consider the following bidifferential operator on $C^\infty(T^*\mathcal{G})$ functions,
\begin{equation}
\Lambda=\lpdv{q^\mu}\rpdv{p_\mu}-\lpdv{p^\mu}\rpdv{q_\mu}
\end{equation}
such that for $C^\infty$ functions, $f(q, p)$ and $g(q, p)$, we have
\begin{eqnarray}
\omega(f\Lambda,g\Lambda)=f\Lambda g=\{f,g\}
\end{eqnarray}
where
\begin{eqnarray}
\{f,g\}=\frac{\partial f}{\partial q^\mu}\frac{\partial g}{\partial p_\mu}-\frac{\partial f}{\partial p^\mu}\frac{\partial g}{\partial q_\mu}.
\end{eqnarray}
It is the poison bracket and $f\Lambda$ and $g\Lambda$ are two vector fields given by $h\Lambda=X_h=-\{h,\}$.\vskip 2pt
The space $T^*\mathcal{G}$ endowed with this symplectic structure is called the phase space, and will be denoted by $\Gamma$. In order to associate the Hilbert space with the phase space $\Gamma$, we will consider the set of complex functions of integrable square, $\phi(q,p)$ in $\Gamma$, such that
\begin{equation}
    \int dpdq\;\phi^\dagger(q,p)\phi(q,p)<\infty
\end{equation}
is a real bilinear form. In this case $\phi(q,p)=\langle q,p|\phi\rangle$ is written with the aid of
\begin{eqnarray}
\int\;dpdq|q,p\rangle\langle q,p|=1,
\end{eqnarray}
where $\langle \phi|$ is the dual vector of $|\phi\rangle$. This symplectic Hilbert space is denoted by $H(\Gamma)$.
\section{Symplectic Quantum Mechanics and the Galilei Group}\label{Symplectic Quantum Mechanics and the Galilei Group}
In this section, we will study the Galilei group, considering $H(\Gamma)$ as the space of representation. To do so,  consider the unit transformations $U\text{:}\mathcal{H}(\Gamma)\rightarrow\mathcal{H}(\Gamma)$ such that $\langle \psi_1|\psi_2\rangle$ is invariant.
Using the $\Lambda$ operator, we define a mapping $e^{i\frac{\Lambda}{2}}=\star\text{:}\Gamma\times\Gamma\rightarrow\Gamma$ called as Moyal (or star) product, defined by.
\begin{eqnarray*}
f\star g=f(q,p)\text{exp}\left[\frac{i}{2}\left(\lpdv{q^\mu}\rpdv{p_\mu}-\lpdv{p^\mu}\rpdv{q_\mu}\right)\right]g(q,p),\nonumber\\
\end{eqnarray*}
it should be noted that we used $\hbar=1$. The generators of U can be introduced by the following (Moyal-Weyl) star-operators:
\begin{equation*}
\widehat{F}=f(q,p)\star=f\left(q^\mu+\frac{i}{2}\frac{\partial}{\partial p_\mu},p^\mu-\frac{i}{2}\frac{\partial}{\partial q_\mu}\right).
\end{equation*}
To construct a representation of Galilei algebra in $\mathcal{H}$, we define the following operators,
\begin{subequations}
\begin{eqnarray}
\widehat{P}^\mu&=&p^\mu\star=p^\mu-\frac{i}{2}\frac{\partial}{\partial q_\mu},\label{eq__p1}\\
\nonumber\\
\widehat{Q}^\mu&=&q\star=q^\mu+\frac{i}{2}\frac{\partial}{\partial p_\mu}.
\end{eqnarray}
and\nonumber
\begin{eqnarray}
\widehat{M}_{\nu\sigma}&=&M_{\nu\sigma}\star=\widehat{Q}_\nu\widehat{P}_\sigma-\widehat{Q}_\sigma\widehat{P}_\nu.
\end{eqnarray}
\end{subequations}
Where $\widehat{M}_{\nu\sigma}$ are the generators of homogeneos transformations and $\widehat{P}_\mu$ of the non-homogeneous. From this set of unitary operators we obtain, after some simple calculations, the following set of commutations relations,
\begin{eqnarray*}
\left[\widehat{P}_\mu, \widehat{M}_{\rho\sigma}\right]&=&-i(g_{\mu\rho}\widehat{P}^\sigma-g_{\mu\sigma}\widehat{P}^\rho),\\
\nonumber\\
\left[\widehat{P}_\mu, \widehat{P}_{\sigma}\right]&=&0,
\end{eqnarray*}
and
\begin{eqnarray*}
\left[\widehat{M}_{\mu\nu},\widehat{M}_{\rho\sigma}\right]=-i(g_{\nu\rho}\widehat{M}_{\mu\sigma}-g_{\mu\rho}\widehat{M}_{\nu\sigma}+g_{\mu\sigma}\widehat{M}_{\nu\rho}-g_{\mu\sigma}\widehat{M}_{\nu\rho}).
\end{eqnarray*}
Consider a vector $q^\mu\in G$ that obeys the set of linear transformations of the type
\begin{eqnarray}
\bar{q}^\mu=G^\mu_{\text{ \;}\nu} q^\nu+a^\mu.\label{translin}
\end{eqnarray}
A particular case of interest of these transformation, given by
\begin{eqnarray}
      \bar{q}^i &=& R^i_jq^j+v^iq^4+ a^i\label{G-1}\\
       \bar{q}^4 &=& q^4+a^4\label{G-2}\\
       \bar{q}^5&=& q^5-(R^i_jq^j)v_i+\frac{1}{2}\textbf{v}^{2}q^4.\label{G-3}
\end{eqnarray}
In the matrix form, the homogeneous transformations are written as

\begin{eqnarray}
G^\mu_{\text{ \;}\nu}=
 \left(\begin{array}{ccccc}
   R^1_{\text{\;}1}&R^1_{\text{\;}2}&R^1_{\text{\;}3}&v^i&0\\
   R^2_{\text{\;}1}&R^2_{\text{\;}2}&R^2_{\text{\;}3}&v^2&0\\
   R^3_{\text{\;}1}&R^3_{\text{\;}2}&R^3_{\text{\;}3}&v^3&0\\
   0&0&0&1&0\\
   v_iR^i_{\text{\;}j}&v_iR^i_{\text{\;}2}&v_iR^i_{\text{\;}3}&\frac{\textbf{v}^2}{2}&1
   \end{array}\right).
\end{eqnarray}
We can write the generators as
\begin{eqnarray}
\begin{aligned}[c]
\widehat{J}_i&=\frac{1}{2}\epsilon_{ijk}\widehat{M}_{jk},\\
\widehat{K}_i&=\widehat{M}_{5i},
\end{aligned}
\qquad
\begin{aligned}[c]
\widehat{C}_i&=\widehat{M}_{4i},\\
\widehat{D}&=\widehat{M}_{54}.
\end{aligned}
\end{eqnarray}
Hence, the non-vanishing commutation relations can be rewritten as
\begin{eqnarray}
\begin{aligned}[c]
\left[\widehat{J}_i,\widehat{J}_j\right]&=i\epsilon_{ijk}\widehat{J}_k,\\
\left[\widehat{J}_i,\widehat{C}_j\right]&=i\epsilon_{ijk}\widehat{C}_k,\\
\left[\widehat{D},\widehat{K}_i\right]&=i\widehat{K}_i,\\
\left[\widehat{P}_4,\widehat{D}\right]&=i\widehat{P}_4,\\
\left[\widehat{P}_i,\widehat{K}_j\right]&=i\delta_{ij}\widehat{P}_5,\\
\left[\widehat{P}_4,\widehat{K}_i\right]&=i\widehat{P}_i,\\
\left[\widehat{D},\widehat{P}_5\right]&=i\widehat{P}_5,
\end{aligned}
\quad
\begin{aligned}[c]
\left[\widehat{J}_i,\widehat{K}_j\right]&=i\epsilon_{ijk}\widehat{K}_k,\\
\left[\widehat{K}_i,\widehat{C}_j\right]&=i\delta_{ij}\widehat{D}+i\epsilon_{ijk}J_k,\\
\left[\widehat{C}_i,\widehat{D}\right]&=i\widehat{C}_i,\\
\left[\widehat{J}_i,\widehat{P}_j\right]&=i\epsilon_{ijk}\widehat{P}_k,\\
\left[\widehat{P}_i,\widehat{C}_j\right]&=i\delta_{ij}\widehat{P}_4,\\
\left[\widehat{P}_5,\widehat{C}_i\right]&=i\widehat{P}_i.\\
\end{aligned}
\end{eqnarray}
This relations form a subalgebra of the Lie algebra of Galilei group in the case of $\mathcal{R}^3\times \mathcal{R}$, considering $J_i$ the generators of rotations $K_i$ and $C_i$ of the pure Galilei transformations, $P_\mu$ the spacial and temporal translations and $D$ of the kind temporal dilatation (which we will not discuss here). In fact, we can observe that eqs. $\eqref{G-1}$ and
$\eqref{G-2}$ are the Galilei transformations given by eq. $\eqref{tg-1}$ and $\eqref{tg-1}$, with $x^4=t$. The eq. $equationre{G-3}$ is the compatibility condition which represents the embedding $$\mathcal{I}: \textbf{A}\rightarrow A=\left(\textbf{A},A_4, \frac{\textbf{A}^2}{2A_4}\right);\quad\textbf{A}\in \mathcal{E}_3, A\in \mathcal{G}$$. The commutation of $K_i$ and $P_i$ is naturally non-zero in this context, so $P_5$ will be related with mass. Which is the extension parameter of the Galilei group. That is an invariant of the extended Galilei-Lie algebra. So the invariants of this algebra in  light cone of de Sitter space-time are
\begin{eqnarray}
I_1&=&\widehat{P}_\mu \widehat{P}^\mu\label{I-1}\\
I_2&=&\widehat{P}_5=-mI\label{I-2}\\
I_3&=&\widehat{W}_{5\mu}\widehat{W}^\mu_5\label{I-3}.
\end{eqnarray}
Where $I$ is the identity operator, $m$ is the mass, $W_{\mu\nu}=\frac{1}{2}\epsilon_{\mu\alpha\beta\rho\nu}P^\alpha M^{\beta\rho}$ is the 5-dimensional Pauli-Lubanski tensor, and $\epsilon_{\mu\nu\alpha\beta\rho}$ is the totally anti-symmetric tensor in five dimensions. In the scalar represantation we can defined $I2=0$.
Using the Casimir invariants $ I_1 $ and $ I_2 $ and applying in $\Psi$, we have:
\begin{eqnarray*}\label{DKP-22}
   \begin{array}{ll}
     \widehat{P}_{\mu}\widehat{P}^{\mu}\Psi=k^2\Psi\label{eq:g1}\\
      \widehat{P}_{5}\Psi=-m\Psi
       \end{array}
\end{eqnarray*}
We obtain
\begin{eqnarray*}
   \left(p^2-i\text{p}\cdot\nabla-\frac{1}{4}\nabla^2-k^2\right)\Psi=2\left(p_4-\frac{i}{2}\del_t\right)\left(p_5 -\frac{i}{2}\del_5\right)\Psi,
\end{eqnarray*}
a solution for this equation is
\begin{eqnarray}
\Psi=e^{-i2p_5q^5}\rho(q^5)e^{-2ip_4t}\chi(t)\Phi(\textbf{q},\textbf{p}).
\end{eqnarray}
Thus,
\begin{eqnarray*}
   \left(p^2\Phi-i\text{p}\cdot\nabla\Phi-\frac{1}{4}\nabla^2\Phi-k^2\right)\frac{1}{\Phi}=\frac{1}{2}\left(i\del_t\chi\right)\left(i\del_5\rho\right)\frac{1}{\chi\rho},
\end{eqnarray*}
this yelds
\begin{eqnarray*}
i\del_t\chi=\alpha\chi,\text{ and }i\del_5\rho=\beta\rho.
\end{eqnarray*}
Thus, our solution for $\chi$ and $\rho$ is
\begin{eqnarray}
\chi=e^{-i\alpha t},\quad \rho=e^{-i\beta q^5}.
\end{eqnarray}
Using the fact that
\begin{eqnarray*}
\widehat{P}_4\Psi=(p_4-\frac{i}{2}\del_t)e^{-i(2p_4+\alpha)t}=-E\,e^{-i(2p_4+\alpha)t},
\end{eqnarray*}
and
\begin{eqnarray*}
\widehat{P}_5\Psi=(p_5-\frac{i}{2}\del_5)e^{-i(2p_5+\beta)q^5}=-m\,e^{-i(2p_5+\beta)q^5}.
\end{eqnarray*}
We can conclude that
\begin{eqnarray}
\alpha=2E,\quad \beta=2m.
\end{eqnarray}
So, we have
\begin{equation*}
    \frac{1}{2m}\left(p^2-i\boldsymbol{p}\cdot\nabla-\frac{1}{4}\nabla^2\right)\Phi=\Big(E+\frac{k^2}{2m}\Big)\Phi,
\end{equation*}
which is the usual form of the Schr\"{o}dinger equation in the phase space for the free particle with mass $m$, with an additional kinect energy of $\frac{k^2}{2m}$, that we can always set as the zero of energy.\vskip 2pt
This equation, and its complex conjugate, can also be obtained by the Lagrangian density in phase space (we use $d^\mu=d/dq_\mu$)
\begin{eqnarray*}
    \mathcal{L}&=&\del^\mu\Psi(q,p)\del\Psi^*(q,p)+\frac{i}{2}p^\mu[\Psi(q,p)\del^\mu\Psi^*(q,p)\\
    &-&\Psi^*(q,p)\del^\mu\Psi(q,p)]+\left[\frac{p^\mu p_\mu}{4}-k^2\right]\Psi\Psi^*.
\end{eqnarray*}
The association of this representation with the Wigner formalism is given by
\begin{eqnarray*}
f_w(q,p)=\Psi(q,p)\star\Psi^\dagger(q,p)
\end{eqnarray*}
where $f_w(q,p)$ is the Wigner function. To prove this, we recall
that the eq. \eqref{eq:g1} can be written as
\begin{eqnarray*}
\widehat{P}_{\mu}\widehat{P}^{\mu}\Psi=p^2\star\Psi(q,p).
\end{eqnarray*}
Multiplying the right hand side of the above equation by$\Psi^\dagger$, we obtain
\begin{eqnarray}
(p^2\star\Psi)\star\Psi^\dagger=k^2\Psi\star\Psi^\dagger,\label{eq:mov1}
\end{eqnarray}
but, $\Psi^\dagger\star p^2=k^2\Psi^\dagger$, thus
\begin{eqnarray}
\Psi\star(\Psi^\dagger\star p^2)=k^2\Psi\star\Psi^\dagger\label{eq:mov2}.
\end{eqnarray}
Subtracting \eqref{eq:mov2} from \eqref{eq:mov1}, we have
\begin{eqnarray}
p^2 \star f_w(q,p)-f_w(q,p)\star p^2=0\label{eq:mov3},
\end{eqnarray}
which is the Moyal brakets $\{p^2,f_w\}_M$. From eq. \eqref{eq__p1} the eq. \eqref{eq:mov3} becomes
\begin{eqnarray}
p_\mu\del_{q_\mu}f_w(q,p)=0.\label{W-L}
\end{eqnarray}
Where the Wigner function in the galilean manifold is a solution of this equation.
\section{Spin 1/2 Symplectic Representaion}\label{Spin 1/2 Symplectic Representaion}
\vskip 2pt
In order to study the representations of spin particles 1/2, we will introduce the $\gamma^\mu\widehat{P}_\mu$, where $\widehat{P}_\mu=p_\mu-\frac{i}{2}\partial_\mu$ in such a way that acting on the 5-spinor in the phase space $\Psi(q,p)$, we have
\begin{equation}
     \gamma^\mu\left(p_\mu-\frac{i}{2}\partial_\mu\right)\Psi(p,q)=k\Psi(p,q),\label{eq:DPS}
\end{equation}
which is the galilean covariant Pauli-Schr\"{o}dinger equation. Consequently the mass shell condition is obtained by following the usual steps.
\begin{eqnarray}
(\gamma^\mu\widehat{P}_\mu)(\gamma^\nu\widehat{P}_\nu)\Psi(q,p)=k^2\Psi(q,p),
\end{eqnarray}
therefore
\begin{eqnarray}
\gamma^\mu\gamma^\nu(\widehat{P}_\mu\widehat{P}_\nu)=k^2=\widehat{P}^\mu\widehat{P}_\nu,
\end{eqnarray}
since $\widehat{P}_\mu\widehat{P}_\nu=\widehat{P}_\nu\widehat{P}_\mu$, we have
\begin{eqnarray}
\frac{1}{2}(\gamma^\mu\gamma^\nu+\gamma^\nu\gamma^\mu)\widehat{P}_\mu\widehat{P}_\nu= \widehat{P}^\mu\widehat{P}_\nu,
\end{eqnarray}
so
\begin{eqnarray}
\left\{\gamma^\mu,\gamma^\nu\right\}=2g^{\mu\nu}
\end{eqnarray}
The eq. \eqref{eq:DPS} can  be derive from the Lagrangian density for spin 1/2 particles in phase space, which is given by
\begin{equation*}
\mathcal{L}=-\frac{i}{4}\Big((\del_\mu\bar{\Psi})\gamma^\mu\Psi-\bar{\Psi}(\gamma^\mu\del_\mu\Psi)\Big)-\bar{\Psi}(k-\gamma^\mu p_\mu)\Psi.
\end{equation*}
where $\bar{\Psi}=\zeta\Psi^\dagger$, with$\zeta=-\frac{i}{\sqrt{2}}\{\gamma^4+\gamma^5\}=
\left(\begin{array}{cc}
0&-i\\
i&0
\end{array}\right).$
For the Galilean covariant Pauli-Schr\"{o}dinger equation case, the association to the Wigner function is given by $f_w= \Psi\star\bar{\Psi},$ with each component satisfying eq. \eqref{W-L}.

Let us now examine the gauge symmetries in phase space demanding the invariance of the Lagrangian by a local gauge transformation given by $e^{\Lambda(q, p)}\Psi$. This leads to the minimum coupling,
\begin{equation*}
\widehat{P}_\mu\Psi\rightarrow\left(\widehat{P}_\mu-eA_\mu\right)\Psi=\left(p_\mu-\frac{i}{2}\del_\mu-eA_\mu\right)\Psi,
\end{equation*}
This describes an electron in an external field, with the Pauli-Schr\"{o}dinger equation given by
\begin{eqnarray}
\left[\gamma^\mu\left(p_\mu-\frac{i}{2}\del_\mu-eA_\mu\right)-k\right]\Psi=0.\label{eq:D-S-F}
\end{eqnarray}
In order to illustrate such result, lets consider a electron in a external field given by $A_\mu(\textbf{A},A_4,A_5)$, with $A_4=-\phi$ and $A_5=0$. Considering the following represantation of $\gamma^\mu$
matrices
\begin{eqnarray*}
\gamma^i=
\left(\begin{array}{cc}
\sigma^i&0\\
0&-\sigma^i
\end{array}\right),\text{ }
\gamma^4=
\left(\begin{array}{cc}
0&0\\
\sqrt{2}&0
\end{array}\right),\text{ }
\gamma^5=
\left(\begin{array}{cc}
0&-\sqrt{2}\\
0&0
\end{array}\right).
\end{eqnarray*}
where $\sigma^i$ are the Pauli matrices and $\sqrt{2}$ is the identity matrix 2x2 multiplied by $\sqrt{2}$.  We can rewrite the object $\Psi$, as
$
\Psi=\left(\begin{array}{cc}
\varphi\\
\chi
\end{array}\right),
$
where $\varphi$ and $\chi$ are 2-spinors dependents dos $x^\mu; \mu=1,...,5$. Thus, in the representation where $k=0$, the eq. \eqref{eq:D-S-F} becomes
\begin{eqnarray}
\boldsymbol{\sigma}\cdot\left(\textbf{p}-\frac{i}{2}\del_q-e\textbf{A}\right)\varphi-\sqrt{2}\left(p_5-\frac{i}{2}\del_5\right)\chi=0,\nonumber\\
\nonumber\\
\sqrt{2}\left(p_4-\frac{i}{2}\del_t-e\phi\right)\varphi-\boldsymbol{\sigma}\cdot\left(\textbf{p}-\frac{i}{2}\del_q-e\textbf{A}\right)\chi=0.\nonumber\\
\label{eq:D-S-F_2}
\end{eqnarray}
Solving the coupled equations  we get an equation for $\varphi$ and $\chi$, and replacing the eigenvalues of $\widehat{P}_4$ and $\widehat{P}_5$, we have
\begin{eqnarray*}
\left[\frac{1}{2m}\left(\boldsymbol{\sigma}\cdot\left(\textbf{p}-\frac{i}{2}\del_q-e\textbf{A}\right)\right)^2+e\phi\right]\varphi&=&E\varphi,\\
\\
\left[\frac{1}{2m}\left(\boldsymbol{\sigma}\cdot\left(\textbf{p}-\frac{i}{2}\del_q-e\textbf{A}\right)\right)^2+e\phi\right]\chi&=&E\chi.\\
\end{eqnarray*}
These are the non-covariant form of the Pauli-Schr\"{o}dinger equations in phase space independent of time, with
\begin{eqnarray*}
f_w=\Psi\star\bar{\Psi}=i\varphi\star\chi^\dagger-i\chi\star\varphi^\dagger,
\end{eqnarray*}
which leads to
\begin{equation*}
E_n=\frac{eB}{m}\Bigg(n+\frac{1}{2}-\frac{s}{2}\Bigg)-\frac{k^2}{2m}
\end{equation*}
where $s=\pm 1$. It should be noted that the above expression represents the Landau levels which shows the spin-splitting feature.
\begin{figure}[!htbp]
\centering
\caption{Wigner Function (cut in $q_1$,$p_1$),Ground State.\\}\label{W-0}
\centering
\begin{tikzpicture}[scale=0.58]
\begin{axis}[point meta min=-0.4, point meta max=0.4, colorbar, colormap/viridis,
    xlabel = $p_1$,
    ylabel = $q_1$,
    view={30}{30}
]
\addplot3[
    surf, faceted
color=cyan, domain=-3:2,y domain=-2:6
]
{(1/pi)*exp(-(x^2+1+(x-y)+1/4*(y^2+1))))};
\end{axis}
\end{tikzpicture}
\end{figure}
\begin{figure}[!htbp]
\centering
\caption{Wigner Function (cut in $q_1$,$p_1$), First Excited State.\\}\label{W-1}
\centering
\begin{tikzpicture}[scale=0.58]
\begin{axis}[point meta min=-0.4, point meta max=0.4, colorbar, colormap/viridis,
    xlabel = $p_1$,
    ylabel = $q_1$,
    view={30}{30}
]
\addplot3[
    surf, faceted
color=cyan, domain=-3:2,y domain=-2:6
]
{1/pi)*((2*(x^2+1+(x-y)+1/4*(y^2+1)))-1)*exp(-(x^2+1+(x-y)+1/4*(y^2+1)))};
\end{axis}
\end{tikzpicture}
\end{figure}
\newpage
\vskip 8pt
The above figures \eqref{W-0} and \eqref{W-1} shows the Wigner functions for the ground and first excited state respectively in the cut $(q_1,p_1)$. These are the same solution known in the literature using the usual Wigner method.
\section{Concluding Remarks}\label{Concluding Remarks}
We study the spin $ 1/2 $ particle equation, the Pauli-Schr\"{o}dinger equation, in the context of Galilean covariance, considering a symplectic Hilbert space. We begin with a presentation on the Galilean manifold which is used to review the construction of Galilean covariance and the representations of quantum mechanics in this formalism, namely the spin and scalar representation $ 1/2 $, {equation} of Schr\"{o}dinger ({Klein-Gordon}-like) and the Pauli-Schr\"{o}dinger (Dirac-like) {equation} respectively. 

The quantum mechanics formalism in phase space is derived in this context of Galilean covariance giving rise to  representations of spin 0 and spin 1/2 equations. For the spin  1/2 equation, the Dirac-like equation, we study the electron in an external field. With the solution, we were able to recover the non-covariant Pauli-Schr\"{o}dinger equation in phase space and analyse, in this context the Landau levels. 
\vskip3mm
A. \textit{This work was supported by CAPES and CNPq of  Brazil.}
\section*{References} 

\end{document}